\documentclass[12pt,aps,psfig,preprint]{article}
\usepackage{epsf}
\newcommand{\eqb}{\begin{equation}}
\newcommand{\eqe}{\end{equation}}
\newcommand{\dmb}{\begin{displaymath}}
\newcommand{\dme}{\end{displaymath}}
\newcommand{\pd}{\partial}
\newcommand{\ep}{\varepsilon}
\newcommand{\eab}{\begin{eqnarray}}
\newcommand{\eae}{\end{eqnarray}}

\newcommand{\be}{\begin{equation}}
\newcommand{\ee}{\end{equation}}

\begin{document}
\begin{titlepage}
\begin{flushright}
HD-THEP-02-45  \\
UVIC-TH-02-06\\
\end{flushright}
\vspace{0.6cm}

\begin{center}
\Large{{\bf Conformal matter in warped backgrounds}}

\vspace{1cm}

Ralf Hofmann$^{1}$ and Maxim Pospelov$^{2,3}$

\end{center}
\vspace{0.3cm}

\begin{center}
 1. {\em
Institut f\"ur Theoretische Physik\\ 
Universit\"at Heidelberg\\ 
Philosophenweg 16, 69120 Heidelberg, Germany}\vspace{0.5cm}

 2.{\em Department of Physics and Astronomy,
University of Victoria,
     Victoria, BC, V8P 1A1 Canada }\vspace{0.5cm}

3.{\em  Centre for Theoretical Physics, University of Sussex, Brighton
BN1 9QJ,~~UK} 

\end{center}
\vspace{0.5cm}

\begin{abstract}

Warped backgrounds in five dimensional models can provide solutions to 
various hierarchy problems in particle physics if the 
standard model matter is associated with the zero modes of bulk fields with 
nontrivial profiles along the extra dimension.  
We investigate the case of a {\sl conformally} coupled massless scalar.
This field possesses a 
zero mode whose action density is exponentially enhanced along the transverse direction 
in a warped background, {\em i.e.} near the negative tension brane in the RS setting. 
At the zero mode level effective 5D composites of this scalar and its derivatives, 
which are suppressed by powers
of the fundamental gravitational scale, yield 4D interactions of electroweak strength. 
This suggests a certain importance of conformally coupled scalar fields for extra-dimensional 
model building.

\end{abstract} 

\end{titlepage}

\section{Introduction}

The last few years have witnessed an explosion of interest in 
field theoretic models involving extra 
dimensions of spatial signature. 
For a large part this rapid development 
was stimulated by the work of Ref.\,\cite{ADD} where the presence of a small 
number of compact extra dimensions, accessible only to gravity, 
was proposed as a mechanism for lowering the $4+n$-dimensional fundamental scale 
relative to the $4$-dimensional Planck scale. 
In this scenario, the large hierarchy between the electroweak and 
the Planck scale can be re-expressed 
as a large hierarchy between the inverse compactification radius 
and the electroweak scale, and the latter is identified with 
the higher dimensional gravitational scale. 
Another important impedus to extradimensional model building was 
launched by the orbifold construction of Randall and Sundrum (RS) \cite{RS} where the possibility of
a purely geometrical solution to the gauge hierarchy problem was demonstrated. 
The key observation in \cite{RS} is that a conversion of the 5D Planck scale to 
the 4D electroweak scale occurs if a single compact extra dimension is 
sufficiently strongly warped due to the presence of a 
bulk cosmological constant and if the relevant matter is 
localized on a brane with negative tension which is positioned 
at the minimum of the warp factor. 

For both extradimensional approaches to be viable and plausible there is the 
need to (i) stabilize the extra dimension(s) and (ii) 
localize the low-energy Standard Model matter fields on 4D hypersurfaces
in a dynamical fashion. Condition (i) is dictated by the necessity 
to be in agreement with the observed 4D late time cosmology \cite{cosmED}. 
There have been many proposals for a stabilization 
of the set-up in \cite{RS} which by itself does not generate a potential for 
the radion field. General citeria for the 
stabilization of RS-like scenarios were worked out recently in Ref.\,\cite{KOP2002}. 
The most popular scenario uses a minimally coupled bulk scalar field which is 
subject to brane potentials \cite{GoldbergerWise}. Operating in the gravitational 
background of the RS solution, in this scenario a radion potential with an 
acceptable minimum is generated by a nontrivial 
bulk profile of the scalar field. Stabilization of extra dimension(s)
due to the Casimir force of conformally coupled bulk matter 
was investigated in Refs.\,\cite{Casimir}. 

In the framework of field theory in Minkowski space 
localization mechanisms for low-energy matter fields are known 
for a long time \cite{JR}. In a realization of (ii) for the warped case the 
weak/Planck hierarchy would appear due 
to the zero mode property of a {\sl dynamically} localized profile in a gravitational 
background which is more fundamental than the framework that simply assumes 
localization \cite{RS}. 

The challenge in the RS framework is to 
localize the zero modes of bulk matter gravitationally. 
A considerable amount of work has been devoted to address 
this question in the past \cite{bulkloc}. It was found 
that the gravitational localization of the massless fermion fields 
is possible while for the gauge fields this mechanism does not work.
 
In case of the gravitational localization of fermions, the 
transverse action density behaves as $a^{-1}$ where $a=a(y)$ denotes 
the warp factor \cite{bulkloc}. This scaling corresponds to 
the strong enhancement near the negative tension brane and can be interpreted as 
localization of the zero mode on the "visible" brane in the RS set-up. 
In case of the minimally coupled bulk scalar field, its zero mode 
has a constant profile along the extra dimension which implies that 
the transverse action density is proportional to $\sim a^2$. 
Thus, the zero mode is effectively localized on the "hidden" brane. 
A nontrivial $y$ profile for the zero mode of a minimally coupled scalar
requires the introduction of extremely fine-tuned sources on 
the branes (see section 2).  
This problem does not arise in the fermionic case since the bulk equation 
of motion is first order.

One may wonder whether a localization of fermionic and scalar 
low-energy matter on different branes is 
an artefact of the minimal coupling of the latter to gravity. 
The main purpose of this Letter is to point out that this is indeed the case. 
We investigate the profiles of the scalar fields 
in curved 5D backgrounds characterized by an arbitrary scale factor $a(y)$. 
We compare the minimally and the conformally coupled cases. 
In the former case we show that nontrivial profiles 
for a zero mode require the introduction of sources on the branes.  De-tuning the 
sources would result in a 
mass of the would-be zero mode of the minimally coupled scalar. 
If the energy scale 
associated with the inverse size of the extra dimension is large, like in the RS model, 
such a mode would not be useful for low-energy phenomenology. 

We then address the case of the conformally coupled scalar. In this case 
we show how one can construct a nontrivial zero mode profile which (i) 
automatically solves the associated scalar field equation and scales as $\sim a^{-3/2}$, (ii) 
does not require a fine-tuning of brane sources to 
ensures the masslessness condition
, and (iii) in the case of a large 
warping exhibits a localization of 
the transverse action density on the negative tension brane. 
In a next step we investigate how interactions between 
conformally coupled scalars in the bulk reduce to the 
interaction between the zero modes in the effective 4D theory. We argue that effective 
terms in the bulk potential, which are suppressed by the inverse powers 
of the Planck scale, in the 4D theory are enhanced due to the $\sim a^{-3/2}$ scaling 
and become electroweak size interactions. 
 
We believe that these features are worth being advocated in view of an 
ample use of conformally coupled scalars in extra-dimensional
model building. Moreover, these scalars can be thought of as being composites of 
gauge and/or fermionic fundamentals that when minimally 
are also conformally coupled. Prime examples for such composites 
are the Goldstone bosons arising from the spontaneous breakdown of a 
continuous, global symmetry. 

\section{Minimally coupled scalar}

Throughout this paper we assume that the geometry in the transverse $y$ direction 
is given by a scale factor $a=a(y)$,
\be
\label{RS}
ds^2 = a^2(y) \eta_{\mu\nu}x^\mu x^\nu  + dy^2\,
\ee
where $\eta_{\mu\nu}\equiv\mbox{diag}(-1,1,1,1)$. 
We assume that the extra dimension is compact and 
has two special points at $y=0$ and $y=L$ according to the 
standard $S_1/Z_2$ orbifold hypothesis. Moreover, 
we assume that $a(y)$ and a massless, noninteracting 
minimally coupled scalar $\phi$ in the bulk are 
even under the orbifold reflection $y\rightarrow -y$. 3-branes 
and sources $V_{0,L}$ for $\phi$ on these branes can be placed at the fixed 
points $y=0,L$. For most of what follows the precise 
form of the given scale factor $a(y)$ is irrelevant. 

We consider the following action for a minimally coupled 
massless bulk scalar field $\phi$ in the background (\ref{RS}),
\eab
\label{minimal}
S = - \int d^{4}x\int dy\sqrt{-g}\left(\frac{1}{2}\,\pd_M{\phi}\,\pd^M{\phi}+
V_0(\phi)\,\delta(y)+V_L(\phi)\,\delta(y-L) \right)\,.
\eae
We are interested in the lowest mass mode in a KK decomposition of $\phi$. 
If the mass of the lowest mode is zero or small compared to other scales in the problem,
the corresponding equation of motion for the $y$ dependence\footnote{In the following 
we denote the zero mode as well as the full field by $\phi$.} in the bulk reads
\be
\label{eomm}
1/a^4 \partial_y (a^4(y) \partial_y \phi(y) ) = 0\,.
\ee
The field $\phi(y)$ is continuous at the boundaries and 
subject to the two boundary conditions, 
\be
\label{jumpm}
\left[\frac{d\phi}{dy}\right](y = 0)= \frac{dV_0}{d\phi}\,,   \;\;\;\;\;\;\;\;\;\;
\left[\frac{d\phi}{dy}\right](y = L)= - \frac{dV_L}{d\phi}\,.
\ee
The square
bracket in (\ref{jumpm}) denotes the discontinuity of its argument and is defined as
\dmb
[f](y)\equiv\lim_{\ep\to 0}\left(f(y+\ep)-f(y-\ep)\right)\,.
\dme 
Eqs.\,(\ref{jumpm}) are usually called jump conditions. 
The general solution for the bulk equation (\ref{eomm}) is \cite{KOP2000},
\be
\phi = B + A\left|\int^{y}_{0} \frac{d{\rm y}}{a^4({\rm y})}\right|\,.
\label{phimin}
\ee
Both $A$ and $B$ are smooth functions of the 4D coordinates $\{x^\mu\}$. 
A nontrivial $y$ dependence of the zero mode, corresponding to $A\not=0$, 
leads to nonvanishing jumps on the branes, $[\pd_y\phi](y=0,L)=2 A/a^4(y=0,L)$, 
and thus needs brane sources according to (\ref{jumpm}). 

A linear combination of $A(x)$ and $B(x)$ is
a massless mode in 4D if it satisfies 
the four dimensional equation of motion 
$\partial_\mu\partial^\mu (B + cA)=0$ and allows for an   
arbitrary rescaling
$ B + cA \rightarrow {\rm const}\times (B + cA)$. 
However, for generic brane potentials $V_{0,L}$ the 
two boundary conditions (\ref{jumpm}) determine $A(x)$ and $B(x)$ algebraically
which means that their magnitudes are fixed and thus 
no $x$ dependence is allowed for. Therefore, none of the solutions with a nontrivial 
profile can represent an effective scalar field with zero mass in 4D.  
A trivial alternative is possible, $A= V_0= V_L=0$ \cite{bulkloc}. 
This leaves $B(x)$ unconstrained, and thus it represents a 
true zero mode. In the RS background the action density for this mode is enhanced 
near the positive tension brane and therefore it is not suitable for model building purposes.  
Another possibility is to introduce an explicit correlation between the two sources $V_{0,L}$ 
and the scale factor $a(y=0,L)$ such that the two boundary 
conditions on $A$ and $B$ degenerate into a single 
one. This, however, is an extreme fine-tuning. 

\section{Conformally coupled scalar}
$\mbox{}$\vspace{0.1cm}\\ 
{\large\sl Field equation and solution}
\vspace{0.5cm}\\ 
We now turn to the case of a conformally coupled bulk scalar in $D$ dimensions.
The corresponding action reads 
\eab
\label{act}
S&=&-\int d^{D}x\sqrt{-g}\left(\frac{1}{2}\,\pd_M{\phi}\,\pd^M{\phi}+
\frac{1}{2}\,\xi_D R_D{\phi}^2 \right)\,,
\eae
where $\xi_D$ is given by $\xi_D\equiv\frac{1}{4}\frac{D-2}{D-1}$. 
One may also add brane-localized sources for the $\phi$ field as in Eq. (\ref{minimal}). 
In conformal coordinates, $ds^2=a^2(z)\eta_{MN}dx^M dx^N$, 
a solution to the scalar field equation in the bulk,
\eqb 
\label{eomcc}
{\phi}^{;M}_{;M}-\xi R{\phi}=0\,,
\eqe
can be obtained by rescaling a solution of the equation of motion in 
a flat background 
\eqb
\label{feom}
\pd_M\pd^M\tilde{\phi}=0\,
\eqe
as ${\phi}=a^{(2-D)/2}(z)\tilde{\phi}$. Here $z$ refers to the 
$D$th coordinate in the {\sl conformal} frame. The most general 
solution $\tilde{\phi}=\tilde{\phi}(z)$ to Eq.\,(\ref{feom}) is
\eqb
\tilde{\phi}(z)=Az+B\,.
\eqe
A finite transformation from RS to conformal coordinates is given as 
\eqb
z(y)=\int_0^y \frac{d{\rm y}}{a(z({\rm y}))}\,.
\eqe
Even orbifold symmetry, $a(z(y))=a(z(-y))$, implies even orbifold symmetry of 
\eqb
\label{phisol}
\phi=a^{(2-D)/2}(z(y))\left\{A\,
\left|\int_0^y \frac{d{\rm y}}{a(z({\rm y}))}\right|+
B\right\}\,.
\eqe
Again, $A$ and $B$ are smooth functions of the 4D coordinates $\{x^\mu\}$ 
and satisfy $\pd_\mu\pd^\mu A=\pd_\mu\pd^\mu B=0$. 
The field $\phi(z(y),\{x^\mu\})$ in Eq.\,(\ref{phisol}) 
is a solution of (\ref{eomcc}) in the RS frame. Note the similarity between (\ref{phimin}) and 
(\ref{phisol}) for $D=5$. However, in contrast to the solution (\ref{phimin}) 
{\sl both} contributions in (\ref{phisol}) have a nontrivial $y$ dependence because of 
an overall $a^{-3/2}$ factor.
\vspace{0.5cm}\\ 
{\large\sl Jump conditions}
\vspace{0.5cm}\\ 
For the discussion of the jump conditions we retreat to $D=5$. 
Since both terms in (\ref{phisol}) have a $y$ dependence the jump 
conditions are expected to severely constrain the bulk solution (\ref{phisol}). 
In RS coordinates we derive the jump conditions for $\phi$ on the branes  
from the scalar field equation that is modified due to the presence of the term 
$\sim\xi_5\phi^2R_5$ in the action: 
\eqb
\label{eomRS}
1/a^4\pd_y\left(a^4\pd_y\phi\right)=\xi_5 R_5\phi\,.
\eqe
The jump conditions become
\eqb
\label{jump}
\frac{[\pd_y\phi](y=0,L)}{\phi(y=0,L)}=
-8 \xi_5 \frac{[\pd_y a](y=0,L)}{a(y=0,L)}\,.
\eqe
The occurence of $[\pd_y a](y=0,L)$ on the RHS of (\ref{jump}) 
originates from a term proportional to $\pd_y^2 a$ in $R_5$. 
Inserting Eq.\,(\ref{phisol}) for $D=5$ into Eq.\,(\ref{jump}) 
and considering the jump at $y=0$, we obtain
\eqb
\label{jc}
(8\xi_5-3/2)\frac{[\pd_y a](y=0)}{a(y=0)}=\frac{1}{a(y=0)}\frac{A}{B}\,.
\eqe
The LHS of (\ref{jc}) vanishes since $8\xi_5=3/2$ and 
consequently we must set $A=0$. For the remaining part
\eqb
\label{visol}
\phi=a^{-3/2}(y)\,B\,
\eqe
the jump conditions (\ref{jump}) 
are {\sl identically} satisfied at $y=0,L$. Unlike the minimally coupled case the nontrivial profile
(\ref{visol}) solves the scalar sector in the background $a(y)$ exactly without the need to 
introduce fine-tuned potentials on the branes.    
 
\section{Zero mode effective 4D Lagrangian}

We now derive a 4D effective action for the associated canonically normalized 4D scalar field $\chi$. The field
$\chi(\{x^\mu\})$ is defined as 
\be
\phi = \frac{B(x)}{a^{3/2}(y)} \equiv \frac{Z^{1/2}\chi(x)}{a^{3/2}(y)}\,.
\ee
Integrating the bulk action density in Eq.\,(\ref{act}) over $y$ in RS coordinates, we obtain 
\be
\label{kinred}
S = -\int_{0}^{L} dy\, \frac{a^4}{a^2}\frac{Z}{a^3}\int d^4x \sqrt{-g_4}\, \partial_\mu \chi 
\partial^\mu \chi\,.
\ee
No potential terms are generated\footnote{The situation is different if gravitational back reaction is
considered. In the absence of a genuine stabilization mechanism, which 
enforces the static geometry assumed in this paper, 
the scale factor $a$ would depend on time due to the stress-energy arising from the scalar sector.} 
for $\chi$ since there is an exact cancellation 
between $R\phi^2$ and $(\pd_y\phi)^2$ terms in the integral over $y$. 
From (\ref{kinred}) we read off an expression for $Z$ in terms of the integral over the inverse
of the scale factor:
\be
\label{kinint}
\frac{1}{Z}= 2\int_0^L \frac{dy}{a(y)}\,. 
\ee
For warped geometries with $a(y=0)\gg a(y=L)$,
the integral (\ref{kinint}) is saturated around $a(y=L)$. 
In particular, using the RS relation to solve the gauge hierarchy problem, 
\be
\label{RSscaling}
a(y=0) = 1 ~~~{\rm and}~~~~a(y=L) \sim M_W/M_{\rm Pl},
\ee 
one immediately concludes that 
\be
\label{Zfactor}
Z = \left(2\int_0^L \frac{dy}{a(y)} \right)^{-1}\sim \kappa\, a(y=L) \sim M_W\,,
\ee
where $\kappa\sim M_{\rm Pl}$ denotes the value of the coefficient 
in the RS exponential and $M_W$ denotes the weak scale. It is worth emphasizing 
that the precise functional form of the scale factor $a(y)$ is not 
important for an estimate of $Z$. For example, the $\cosh$-like solutions \cite{cosh} 
with the hierarchy (\ref{RSscaling}) generate a similar value for $Z$. 

On the level of zero modes we 
may now break perturbatively the conformal symmetry of the 
scalar sector by adding higher dimensional
composites to the 5D Lagrangian in Eq.\,(\ref{act}). Using the estimate of the 
$Z$-factor (\ref{Zfactor}) in the RS framework, we then may investigate the 
effective 4D scaling of these operators. Let us first 
look at nonderivative couplings. Terms in the bulk action, that are suppressed
by inverse powers of the UV cutoff parameter $M_{\rm Pl}$, are
\be
\label{Intact}
S_n^{(5)} = -\int d^4x \int dy\, a^4 \phi^n M_{\rm Pl}^{5-\frac{3n}{2}}\,,
\ee 
where $n>3$. The action (\ref{Intact}) can be generalized to the 
case of several conformally coupled scalar fields. When reduced to the zero mode 
level and after integrating over $y$ 
Eq.\,(\ref{Intact}) becomes 
\be
\label{redzer}
S_n^{(4)} = -\int d^4x\, \chi^n Z^{\frac{n}{2}}
M_{\rm Pl}^{4-\frac{3n}{2}}a^{4-\frac{3n}{2}}(y=L)\sim  -\int d^4x\, \chi^n M_W^{4-n}\,.
\label{scaling}
\ee
Eq.\,(\ref{redzer}) shows that at the zero mode level Planck scale suppressed operators in 5D 
induce interactions in the effective 4D theory that are of electroweak strength.
For example, a quartic term $\phi^4$ in the bulk, being 
suppressed by the first power of $M_{\rm Pl}$, 
gives rise to a quartic interaction of the zero modes of strength $\sim 1$. 
It is also instructive to look at the bilinear term $M^2 \phi^2$ where $M$
is some mass parameter. Its scaling 
is different from (\ref{scaling}) because the integral over $y$ is saturated near 
$a_0$. The requirement that the mass term for the zero mode arising from this 
operator should be of order $M_W^2\chi^2$ implies that $M\sim\sqrt{M_{\rm Pl}M_W}$.

A similar picture arises for the $M_{\rm Pl}$-suppressed operators that contain derivatives such as 
\eqb
\label{der}
M_{\rm Pl}^{5-5n}\left(\partial_M\phi\, \partial^M \phi\right)^n\,.
\eqe
These operators are relevant for  
the interaction  of Goldstone bosons. For exponential warping and on 
the zero mode level terms $\sim (\pd_y \phi \partial^y \phi)^n$ 
can be treated as polynomial interactions 
$\sim M_{\rm Pl}^{-2}\phi^{2n}$. It is then easy to see that 
the 5D operator (\ref{der}) induces 
a 4D chain of operators of the form 
\eqb
\label{der4D}
\sim -\int d^4x\,  M_W^{4-4r-2s}(\pd_\mu\chi\pd^\mu\chi)^r\chi^{2s}\,,\ \ \ \ \ (r+s=n)\,.
\eqe
Thus, pure derivative couplings in 5D generate weak-scale suppressed mixed 
derivative and polynomial operators. 

\section{Conclusions}

We have shown that in a 
gravitational background given by a single warp factor 
$a=a(y)$ and in the setting of Ref.\,\cite{RS} a 
conformally coupled bulk scalar field has a zero mode with a non-trivial 
$y$-profile, $\phi \sim a^{-3/2}$. This leads to localization 
of this zero mode at the negative tension brane. The behaviour is similar
to what has been observed for the minimally coupled fermion fields ($\psi \sim a^{-2}$). 
We show that such zero modes can be used in models that 
have an RS type of solution to the gauge hierarchy problem.
In particular, we demonstrate that the effective 5D interactions of 
the zero modes, which are suppressed by powers of the Planck scale, 
correspond to the effective 4D interactions of electroweak 
strength. Similar results should also hold for fermions. The missing 
piece then is, of course, a plausible localization 
mechanism for gauge fields in the RS context.

Apart from the gauge hierarchy problem there are proposals to use nontrivial 
bulk profiles for the resolution of the fermion flavor hierarchy \cite{Shifman} and
vrious cosmological  problems 
\cite{Suss}, etc. We have shown that such a 
nontrivial scalar bulk profile exists ($B=\mbox{const}\neq 0$ in (\ref{visol})) 
in the conformally coupled case 
as an exact solution to the scalar field equation in a warped background. 

The conformal coupling of the scalar 
sector to an AdS background breaks AdS SUSY explicitly. As far as 
grand unification is concerned, it has been advocated recently that a high-scale non-SUSY realization 
in a compactified AdS background is possible due 
to a logarithmic running of the gauge coupling even 
beyond the lowest KK excitation energy \cite{logrun}.

\section*{Acknowledgements}    

One of us (R.H.) would like to thank N. Tetradis for a stimulating conversation and 
Z. Tavartkiladze for various discussions. M.P. is partly supported by the NSERC of Canada and 
PPARC of the UK.

\bibliographystyle{prsty}

\end{document}